\documentclass[aps,pra,epsfigure,twocolumn,showpacs]{revtex4}
\usepackage{dcolumn}
\usepackage{bm}
\usepackage{graphicx}
\usepackage{amsmath}
\usepackage{latexsym}
\usepackage{amsfonts}
\usepackage{amssymb}
\usepackage{array}
\usepackage{epsfig}
\usepackage{times}
\usepackage{txfonts}
\usepackage{epstopdf}
\usepackage{pifont}
\usepackage{dsfont}
\usepackage{amscd}
\usepackage{amsfonts}
\usepackage{textcomp}
\usepackage{float}
\usepackage{mathpazo}

\newcommand{\be}{\begin{equation}}
\newcommand{\ee}{\end{equation}}
\newcommand{\bea}{\begin{eqnarray}}
\newcommand{\eea}{\end{eqnarray}}

\begin{document}
\title{Evaluating entropy rate of laser chaos and shot noise}
\author{Xiaomin Guo$^1$, Tong Liu$^1$, Lijing Wang$^1$, Xin Fang$^1$, Tong Zhao$^1$, Martin Virte$^2$, and Yanqiang Guo$^{1,3,*}$}
\affiliation{$^1$Key Laboratory of Advanced Transducers and Intelligent Control System, Ministry of Education and Shanxi Province, College of Physics and Optoelectronics, Taiyuan University of Technology, Taiyuan 030024, China \\
$^2$Brussels Photonics Team (B-PHOT), Department of Applied Physics and Photonics (TONA), Vrije Universiteit Brussel, Pleinlaan 2, 1050, Brussels, Belgium \\
$^3$State Key Laboratory of Quantum Optics and Quantum Optics Devices, Shanxi University, Taiyuan, 030006, China}
\affiliation{*E-mail: guoyanqiang@tyut.edu.cn}

\begin{abstract}
Evaluating entropy rate of high-dimensional chaos and shot noise from analog raw signals remains elusive and important in information security. We experimentally present an accurate assessment of entropy rate for physical process randomness. The entropy generation of optical-feedback laser chaos and physical randomness limit from shot noise are quantified and unambiguously discriminated using the growth rate of average permutation entropy value in memory time. The permutation entropy difference of filtered laser chaos with varying embedding delay time is investigated experimentally and theoretically. High resolution maps of the entropy difference is observed over the range of the injection-feedback parameter space. We also clarify an inverse relationship between the entropy rate and time delay signature of laser chaos over a wide range of parameters. Compared to the original chaos, the time delay signature is suppressed up to 95\% with the minimum of 0.015 via frequency-band extractor, and the experiment agrees well with the theory. Our system provides a commendable entropy evaluation and source for physical random number generation.

\end{abstract}
\date{\today}
\maketitle

Semiconductor laser with external optical feedback is one of the most prominent system to generate high-dimensional laser chaos and has attracted considerable interest for a range of applications \cite{Soriano13,Sciamanna15}, such as chaotic secure communications \cite{Argyris05,Hong08,Wu11}, physical random number generators \cite{Uchida08,Reidler09,Sunada12}, chaos key distribution \cite{Yoshimura12}, chaotic optical sensing \cite{Lin04,Wang08,Xia13} and computing \cite{Brunner13}. The randomness of laser chaos is fundamental to its security and cryptographic applications. A key challenge arises from the difficulty of quantifying physical process randomness from the raw data. To quantify the randomness of laser chaos, entropy in its many forms, such as Shannon entropy \cite{Mikami12,Chan16}, Kolmogorov-Sinai (KS) entropy \cite{Gaspard98,Boffetta02,Hagerstrom15,Hart17} and permutation entropy (PE) \cite{Bandt02,Toomey14,Li15,Quintero15,Rontani16,Hong17}, provides a way to measure temporal randomness of a physical process. The positivity of the entropy per unit time or entropy rate is an evidence for randomness in time series, which characterizes the production of new and random information of high-dimensional and noisy dynamical systems. The KS entropy is one indicator for measuring the entropy rate, but one needs to construct a very refined partition to support all information on the dynamical system. It is not easy to determine and to estimate from time series measurements of a high-dimensional chaotic or noisy experimental system \cite{Hagerstrom15}. Bandt and Pompe proposed a simple randomness measure (i.e. PE) for time series, based on a comparison of neighboring values \cite{Bandt02}. The advantages of the PE measure reside in its flexibility and simplicity in experimental analysis, fast computation process, and its robustness. It can be applied to any type of time series (regular, chaotic, and noisy), and is particularly used to quantify transition from weak to strong chaos \cite{Toomey14} and complexity \cite{Li15,Quintero15,Hong17}. In chaotic systems, the growth rate of the PE reflects stochastic or noise-dominated properties, which make it almost undistinguishable from ideal noise limit. Meanwhile, laser chaos rapidly and nonlinearly amplifies the uncertainty of intrinsic quantum noise and the chaotic amplification of the intrinsic noise is completely different from output fluctuation caused by deterministic chaos \cite{Sunada12}. It is important to understand the origins of the stochastic properties and assess the randomness of the physical process \cite{Mikami12,Hagerstrom15,Hart17}. Shot noise is an intrinsic noise of chaotic system and a fundamental quantum noise limit of optical source, which represents quantum vacuum state or zero-point fluctuations allowed by the minimum uncertainty product of quantum mechanics \cite{Xiao87,Gabriel10,Guo18}. It is natural to compare and evaluate entropy rates between laser chaos and shot noise. However, characterizing shot noise is mainly through statistical metrics like average rate, variance, and signal-to-noise ratio. Evaluating the entropy rate of high-dimensional chaos and shot noise remains elusive and important in statistical mechanics and information security. In this paper, we experimentally assess the entropy rate of laser chaos and shot noise using the growth rate of average PE value in memory time. Based on homodyne detection and frequency-band extractor, the PE growth of shot noise is achieved for the first time. The PE difference of filtered laser chaos with varying embedding delay time is investigated experimentally and theoretically. The PE difference of extracted shot noise and noise-dominated laser chaos grow faster than linearly with embedding dimension. The entropy rate of the laser chaos and shot noise are unequivocally differentiated. We observe the relationship between the entropy rate and time delay signature (TDS) of laser chaos under different feedback strengths and injection currents-a relationship that, to our knowledge, has not been investigated experimentally. That indicates the entropy rate can reflect the randomness and chaos strength of the system. We also employ the frequency-band extractor to suppress the TDS of laser chaos, and the results are in good agreement with the theoretical simulation. In view of the demonstration, we present a commendable entropy evaluation and source for physical random number generation.

\begin{figure}[htbp]
\centering
\includegraphics[width=\linewidth]{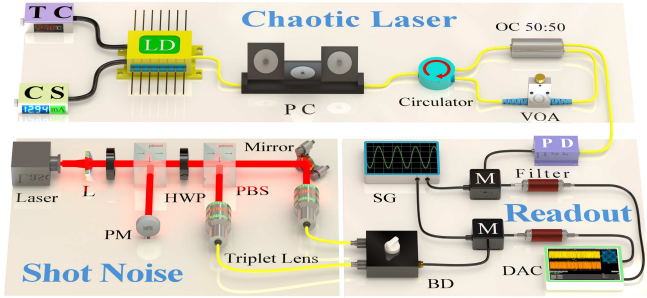}
\caption{Experimental setup. TC, temperature controller; CS, current source; LD, distributed feedback laser diode; PC, polarization controller; VOA, variable optical attenuator; OC, optical coupler; PD, photodetector; M, mixer; SG, signal generator; Filter, low-pass filter; DAS, data acquisition system containing an oscilloscope and a spectrum analyzer; L, lens; PM, power meter; HWP, half-wave plate; PBS, polarizing beamsplitter; BD, balanced-homodyne detector.}
\label{fig1}
\end{figure}

Figure \ref{fig1} depicts the schematic of the experimental layout, which mainly includes two physical entropy sources: a chaotic laser and a shot noise. The chaotic laser consists of a distributed feedback laser diode (LD) subject to optical feedback through a fiber ring cavity. The LD is stabilized at 1550 nm by a temperature controller (TC) and a current source (CS), with the accuracy of 0.01 $^{\circ}$C and 0.1 mA respectively. The free-running threshold current $J_{th}$ is 10.3 mA. The LD is subject to polarization-maintained optical feedback through a fiber loop, and the feedback delay time is 86.7 ns. The VOA with an optical resolution of 0.01 dB and PC are used to accurately control the feedback conditions. The feedback rate $\eta $ is estimated as the ratio of the feedback power to the output power of the LD. From the maximum feedback rate $\eta _{max}=0.5$, the feedback rate $\eta $ could be attenuated by more than 20 dB, with the attenuation defined as $\gamma $ $[dB]=-10\log _{10}(\eta /\eta _{max})$. The output of chaotic laser is acquired through a 50 GHz bandwidth photodetector (PD, Finisar XPDV2120RA). The resulting electrical signal then passes through a frequency-band extractor that is composed of a mixer (M), a low-pass filter (LPF), and a RF signal generator (SG). All data are recorded using a data acquisition system (DAS) which simultaneously measures power spectra by a 26.5 GHz spectrum analyzer (Agilent N9020A, 3 MHz RBW, 3 kHz VBW) and time series by a 36 GHz bandwidth real-time oscilloscope (Lecroy LabMaster10-36Zi) with a maximum sampling rate of 40 GS/s. The length, $10^{6}$, of time series is captured for each time trace.

The shot noise is prepared through homodyne quadrature measurement. A 1550 nm single-mode laser beam is incident on one port of the beamsplitter and acts as the strong local oscillator (LO), while the other port is blocked to ensure that only the quantum vacuum state could enter in. The LO and the vacuum state interfere on a symmetric beamsplitter to form two output beams with balanced powers. A half-wave plate (HWP) and a polarizing beamsplitter (PBS) are combined to serve as accurate 50:50 beam splitting. The two outputs are simultaneously measured with a balanced-homodyne detector (BD, Thorlabs PDB480C). The resulting shot noise signal is extracted by a frequency-band extractor, and the outcome is subsequently fed into the DAS.

\begin{figure}[htbp]
\centering
\includegraphics[width=0.9\linewidth]{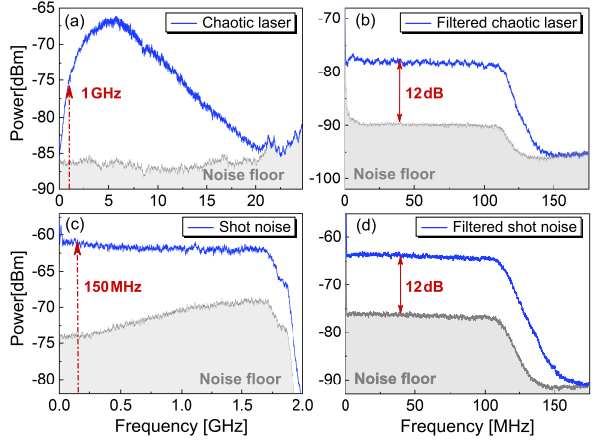}
\caption{(a) Measured original chaotic laser when $J=1.5J_{th}$, $\gamma=4.5$ dB and (c) shot noise power spectrum; (b) filtered chaotic laser and (d) filtered shot noise power spectrum with 100 MHz frequency bandwidth.}
\label{fig2}
\end{figure}

We first extract the same bandwidth spectrum from laser chaos and shot noise to investigate the entropy evolution of physical process. In order to eliminate inter-sample correlations and obtain adequate entropy, the sampling rate used in our experiment approaches twice the signal bandwidth and obeys the Nyquist theorem. Figure \ref{fig2}(a) shows the measured power spectrum of the original chaotic laser operating at injection current $J=1.5J_{th}$ and feedback attenuation of 4.5 dB, and filtered chaotic laser obtained from the frequency-band extraction [Fig. \ref{fig2}(b)]. According to the 80\% bandwidth definition, the bandwidth of the original chaotic laser is about 8.6 GHz. The original chaotic laser is down-mixed using a sinusoidal 1 GHz signal and filtered by a LPF with 100 MHz cutoff frequency (Mini-Circuits BLP100+). It is clear that the filtered chaos has a large clearance above the noise floor of 12 dB. The same clearance of shot noise is achieved by increasing the LO power up to 3 mW, as shown in Fig. \ref{fig2}(c). The resulting shot noise signal is mixed down with a 150 MHz carrier and filtered with a 100 MHz bandwidth LPF [Fig. \ref{fig2}(d)]. The same bandwidth of vacuum sideband frequency spectrum is extracted from the whole measured shot noise.

\begin{figure}[htbp]
\centering
\includegraphics[width=0.9\linewidth]{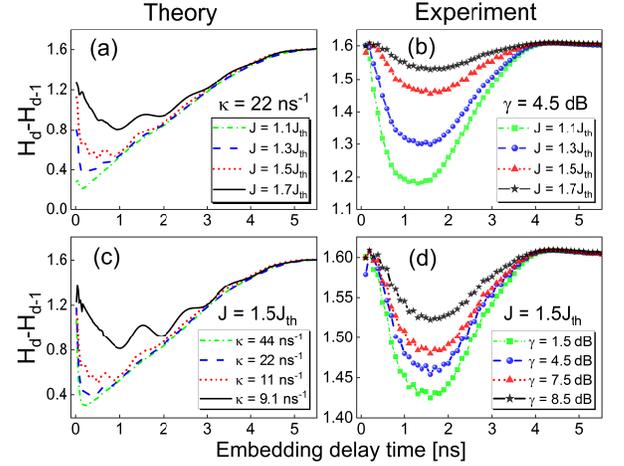}
\caption{(a), (c) Theoretical and (b), (d) experimental results for $H_{d}-H_{d-1}$ of filtered chaotic laser at four $J$ (a), (b): $1.1J_{th}$, $1.3J_{th}$, $1.5J_{th}$, and $1.7J_{th}$; four $\kappa (\gamma )$ (c), (d): $44$ $ns^{-1}$ ($1.5$ dB), $22$ $ns^{-1}$ ($4.5$ dB), $11$ $ns^{-1}$ ($7.5$ dB), $9.1$ $ns^{-1}$ ($8.5$ dB). The embedding dimension $d$ is chosen as 5.}
\label{fig3}
\end{figure}

To investigate the influence of injection current and feedback strength on the entropy evolution of laser chaos, the PE difference $H_{d}-H_{d-1}$ versus embedding delay time $t$ is theoretically modeled by the Lang-Kobayashi (LK) equations \cite{Lang80} and experimentally measured for various $J$ and $\gamma $($\kappa $), as shown in Fig. \ref{fig3}. The LK model is employed to characterize the chaotic dynamics and consists of rate equations for the variables of the complex electric field amplitude $E$ and the carrier density $N$. Parameters are defined in \cite{Lang80}. Namely, $G(t)=G_{N}[N(t)-N_{0}]/(1+\varepsilon \left\vert E(t)\right\vert ^{2})$ is the nonlinear optical gain (with $G_{N}$ the gain coefficient and $\varepsilon $ the saturation coefficient), $N_{0}$ is the carrier density at transparency, $\kappa $ is the feedback strength, $\tau _{p}$ is the photon lifetime, $\tau _{N}$ is the carrier lifetime, $\alpha $ is the linewidth enhancement factor, and $\omega $ is the angular optical frequency. We take $\alpha =5$, $\tau _{p}=3.2$ ps, $\tau _{N}=2.3$ ns, $G_{N}=2.7\times 10^{-8}$ $ps^{-1}$, $N_{0}=1.36\times 10^{8}$, $\omega =1.216\times 10^{15}$ rad/s (i.e., the laser wavelength of 1.55 $\mu $m). The laser threshold current is $J_{th}=(1/\tau_{N})[N_{0}+1/(G_{N}\tau _{p})]\approx 10.3$ mA, corresponding to that of the LD used in the experiment. The external feedback delay time $\tau _{ext}$ is set equal to the experimental value. To mimic the frequency-band extractor, the theoretical time series of chaos are mixed with a sinusoidal signal and filtered using a Chebyshev Type-II LPF with a bandwidth of 100 MHz. The PE definition can be found in Refs. \cite{Bandt02,Toomey14,Li15,Quintero15,Rontani16,Hong17} and the embedding dimension $d$ is chosen as 5 in Fig. \ref{fig3}. At all J and $\gamma $($\kappa $) shown in theory [Fig. \ref{fig3}(a) and \ref{fig3}(c)] and experiment [Fig. \ref{fig3}(b) and \ref{fig3}(d)], the $H_{d}-H_{d-1}$ follows a nonmonotonic dependence on the embedding delay time, revealing a dip that appears for intermediate $t$ corresponding to half of the filtering bandwidth. According to the Nyquist theorem, the maximum $H_{d}-H_{d-1}$ of filtered chaotic laser versus embedding delay time is observed around double filtering bandwidth (i.e., $t=4.7$ ns in the experiment). The theory is in good agreement with the experiment, both in terms of curve shape and its variation with increasing $J$ and $\gamma $. It is also confirmed that the LK equations well model and reproduce the experiment. The dip depth is related to the entropy generation of chaotic process and is also a good indicator of the randomness of chaotic signal. As the $J$ and $\gamma $ increase (or $\kappa $ decreases), the dip depth decreases, i.e. higher injection current or lower feedback strength can induce more entropy production. In other words, the photon and carrier lifetime of chaotic laser have effects on the dip depth, and long photon and carrier lifetime contribute to the dip depth increasing and the entropy growing. It should be noted that the measurement can be transposed to other integrated VCSEL and quantum-dot systems of laser chaos \cite{Virte13,Virte16,Reitzenstein19}.

\begin{figure}[htbp]
\centering
\includegraphics[width=0.8\linewidth]{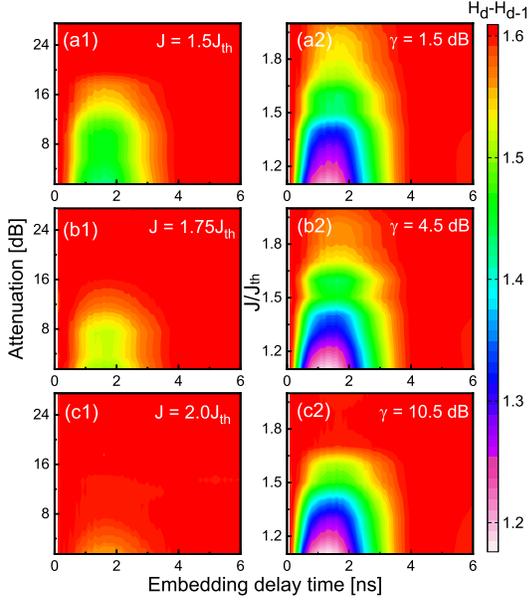}
\caption{Measured maps of the $H_{d}-H_{d-1}$ ($d=5$) of filtered chaotic laser versus embedding delay time with varying injection currents (a1) $J=1.5J_{th}$, (b1) $J=1.75J_{th}$, (c1) $J=2.0J_{th}$ and feedback attenuations (a2) $\gamma =1.5$ dB, (b2) $\gamma =4.5$ dB, (c2) $\gamma =10.5$ dB.}
\label{fig4}
\end{figure}

Under the above acquisition condition, high resolution maps of the $H_{d}-H_{d-1}$ for the filtered laser chaos over the range of the injection-feedback parameter space is further measured in detail. Figure \ref{fig4} shows measured maps of the difference $H_{d}-H_{d-1}$ of the filtered chaotic laser varies with embedding delay time under different injection currents $J$ and feedback attenuations $\gamma $. Here the embedding dimension $d$ is chosen as 5. As can be seen in Fig. \ref{fig4}(a1)-\ref{fig4}(c1), the difference $H_{d}-H_{d-1}$ increases with feedback attenuation $\gamma $ within twice the LPF bandwidth, i.e. the growth is faster than linear and the chaotic dynamics is dominated by noise. As the $J$ decreases, the regions with the variation of $H_{d}-H_{d-1}$ are broadened. For various $\gamma $ [Fig. \ref{fig4}(a2)-\ref{fig4}(c2)], the $H_{d}-H_{d-1}$ grows monotonically with $J$. The regions corresponding to the growth $H_{d}-H_{d-1}$ become larger as $\gamma $ decreases.

\begin{figure}[htbp]
\centering
\includegraphics[width=0.95\linewidth]{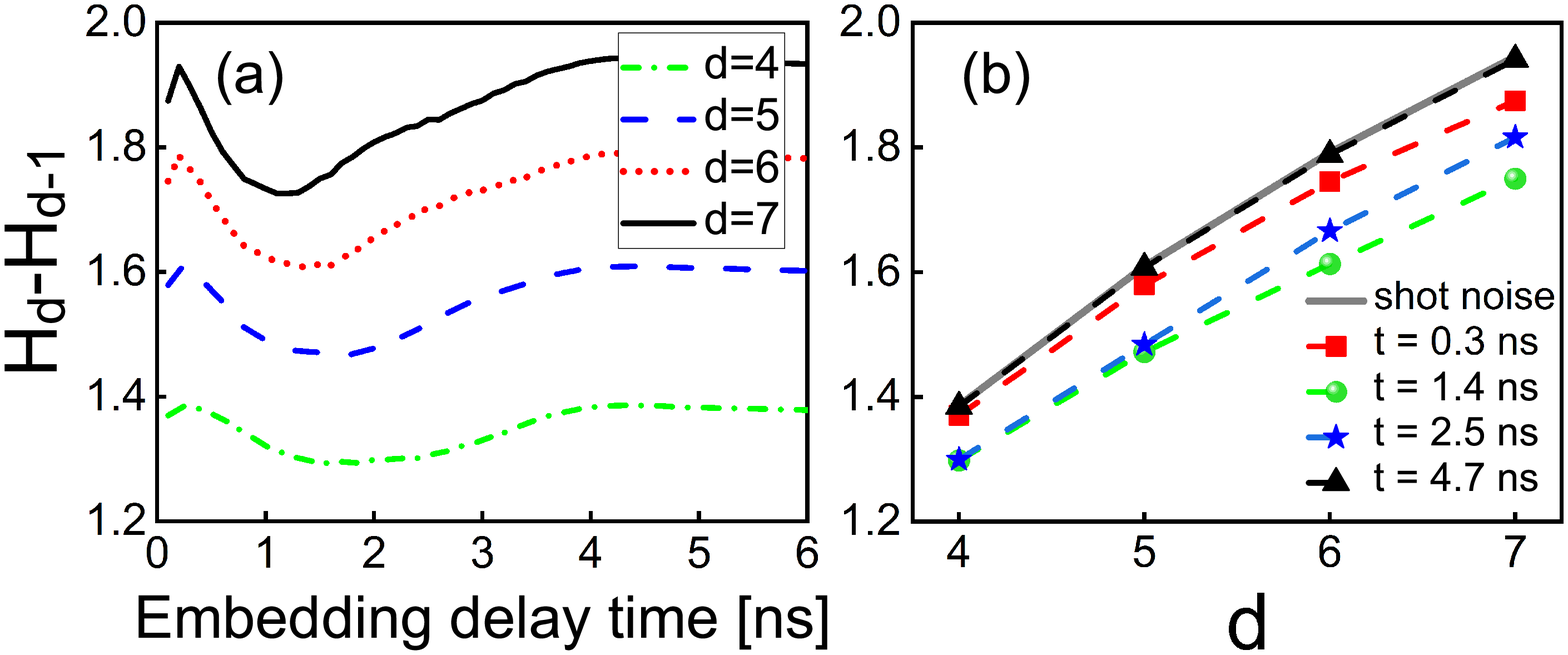}
\caption{Measured difference $H_{d}-H_{d-1}$ of filtered chaotic laser as functions of (a) embedding delay time $t$ and (b) embedding dimension $d$. The shot noise versus $d$ are indicated with the solid gray line.}
\label{fig5}
\end{figure}

To ensure that the dynamical process of filtered chaotic laser and shot noise is dominated by noise, the growth of PE ($H_{d}-H_{d-1}$) with embedding dimension $d$ is measured for various embedding delay time $t$, as shown in Fig. \ref{fig5}. The difference $H_{d}-H_{d-1}$ increases monotonically with $d$. For practical purposes the values of $d$ is used between 4 and 7 in Fig. \ref{fig5}. The injection current and feedback attenuation are the same as in Fig. \ref{fig2}. According to the Nyquist theorem, the maximum $H_{d}-H_{d-1}$ of filtered chaotic laser versus embedding delay time is observed around double filtering bandwidth (i.e., $t=4.7$ ns) in Fig. \ref{fig5}(a). As shown in Fig. \ref{fig5}(b), the $H_{d}-H_{d-1}$ of filtered shot noise (indicated with a solid gray line) and filtered laser chaos at $t=4.7$ ns increase with the increase of $d$, and the values of $H_{d}-H_{d-1}$ is almost consistent with the ideal noise-dominated limit $H_{d}-H_{d-1}=\ln d!-\ln (d-1)!$. The $H_{d}-H_{d-1}$ linear growth with $d$ reveals that the physical process is dominated by noise, but it is noteworthy that the $H_{d}-H_{d-1}$ growth of filtered laser chaos and filtered shot noise can not be discriminated unambiguously as the embedding delay time $t$ increases.

\begin{figure}[htbp]
\centering
\includegraphics[width=0.95\linewidth]{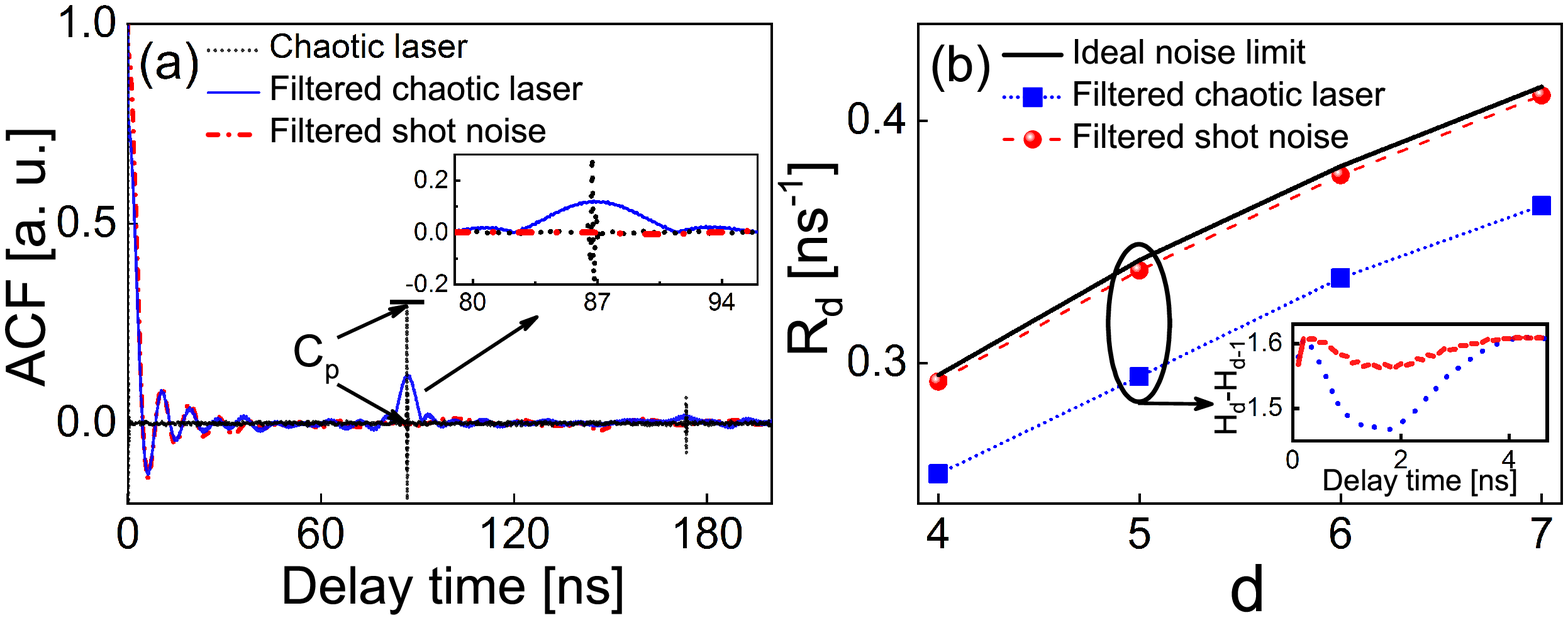}
\caption{(a) ACF of original chaotic laser, filtered shot noise and filtered chaotic laser. The inset shows the peak value $C_{p}$ at feedback delay time of 86.7 ns; (b) Entropy rate $R_{d}$ of filtered shot noise and filtered chaotic laser as a function of the dimension $d$. The entire embedding delay time is $t_{m}=4.7$ ns. The solid line indicates the $R_{d}$ of ideal noise limit. The inset shows the difference $H_{d}-H_{d-1}$ for $d=5$.}
\label{fig6}
\end{figure}

In order to unequivocally differentiate the $H_{d}-H_{d-1}$ growths of filtered laser chaos and filtered shot noise and  further investigate chaotic dynamics and time evolution of entropy, the autocorrelation function (ACF) and entropy rate are unveiled experimentally. Figure \ref{fig6}(a) shows the associated ACF of the original chaotic laser, filtered chaotic laser and filtered shot noise. Parameters are the same as in Fig. \ref{fig2}. The peak value $C_{p}$ of the ACF at the external feedback delay time $\tau _{ext}$ is used to quantify the TDS of chaos \cite{Wu12,Chan15,Cheng15}, which gives useful information on the chaotic dynamics. The existence of the TDS will degrade the randomness of chaos. However, it is obvious that the TDS of the chaotic laser is effectively suppressed (the $C_{p}$ value is reduced from 0.296 to 0.12) by using the frequency-band extractor and the shot noise signal has no redundant correlation or periodicity. To accurately assess the entropy evolution of physical process, we define an entropy rate $R_{d}$ as follows:
\begin{equation}
R_{d}=\frac{\left\langle H_{d}-H_{d-1}\right\rangle }{t_{m}},
\label{eq1}
\end{equation}
where $\left\langle \cdot \right\rangle $ denotes the average over PE differences in $t_{m}$ and $t_{m}$ denotes the memory time corresponding to the Nyquist sample rate. The entropy rate $R_{d}$ of the filtered chaotic laser and filtered shot noise increase with the embedding dimension $d$ as shown in Fig. \ref{fig6}(b). Here the entire embedding delay time is $t_{m}=4.7$ ns that corresponds to the double filtering bandwidth. The entropy rate $R_{d}$ can unambiguously discriminate between filtered chaotic laser and filtered shot noise. The $R_{d}$ of filtered shot noise is very close to the ideal noise limit.

\begin{figure}[htb]
\centering
\includegraphics[width=\linewidth]{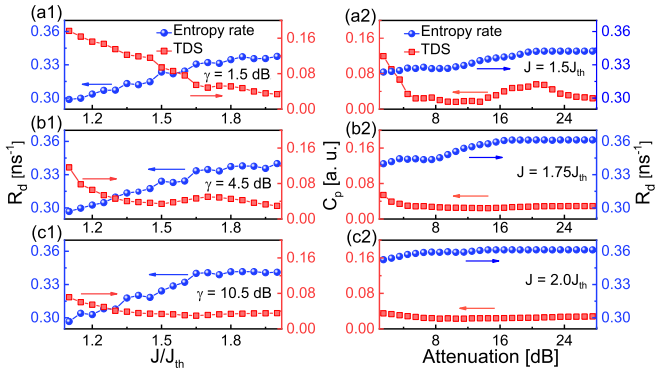}
\caption{Measured entropy rate and TDS of filtered chaotic laser as functions of the injection currents for (a1) $\gamma =1.5$ dB, (b1) $\gamma =4.5$ dB, (c1) $\gamma =10.5$ dB and feedback attenuations for (a2) $J=1.5J_{th}$, (b2) $J=1.75J_{th}$, (c2) $J=2.0J_{th}$, with $d=5$ and $t_{m}=4.7$ ns.}
\label{fig7}
\end{figure}

To characterize the relationship between the entropy rate $R_{d}$ and the TDS $C_{p}$, we experimentally address the effects of injection current and feedback strength on it. Figure \ref{fig7} shows the coexistence of high $R_{d}$ and TDS suppression as functions of the $J$ and $\gamma $, with $d=5$ and $t_{m}=4.7$ ns. At the three $\gamma $ shown [Figs. \ref{fig7}(a1)-\ref{fig7}(c1)], the $R_{d}$ and TDS have an inverse relationship for the injection currents ($1.1J_{th}\leq J\leq 2.0J_{th}$). For small $\gamma $ in Fig. \ref{fig7}(a1), the $R_{d}$ almost linearly increases with increasing injection current and the minimum TDS is observed at the highest injection current being measured. As $\gamma $ increases, the $R_{d}$ increases initially with the increase of $J$ and subsequently shows saturation. After that, the $J$ has little effect on the $R_{d}$ and TDS. The effect of feedback attenuation on the $R_{d}$ and TDS in Figs. \ref{fig7}(a2)-\ref{fig7}(c2) has some similarity with that of the $J$. As $J$ increases, a sharp increase of $R_{d}$ with increasing feedback attenuation appears at the very beginning. Clearly, it is noted that high $R_{d}$ coincides with strong TDS suppression over a wide range of injection currents and feedback attenuations, indicating that the parameters affect the trend of the entropy evolution and chaos strength.

\begin{figure}[htbp]
\centering
\includegraphics[width=\linewidth]{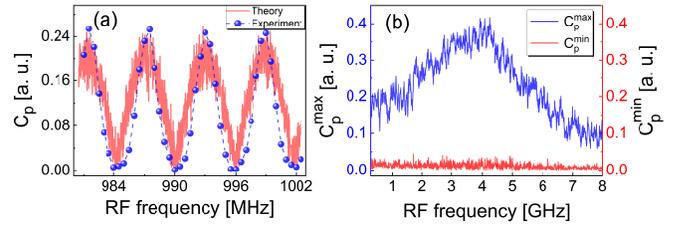}
\caption{(a) Theoretical and experimental TDS of the filtered chaotic laser as a function of RF frequency for $J=1.5J_{th}$ and $\gamma =4.5$ dB ($\kappa =22$ $ns^{-1}$); (b) theoretical results of the maxima and the minima of $C_{p}$ with varying the RF frequency.}
\label{fig8}
\end{figure}

Finally, we also optimize the effect of RF frequency and filtering bandwidth on the TDS suppression and the $R_{d}$ enhancement of the filtered chaotic laser. Figure \ref{fig8}(a) shows experiment and theory for the TDS of filtered chaotic laser under different RFs. The operating parameters of filtered chaotic laser are the same as in Fig. \ref{fig2}, except the RF frequency. The RF signal is generated by the SG, whose frequency is much higher than that of the LPF passband. By adjusting the RF frequency, the TDS of filtered chaotic laser is better reduced to 0.015 compared to the original chaos. This significant reduction varies periodically and the $C_{p}$ values have the minima in every $1/(2\tau _{ext})=6$ MHz. To verify the observation, we employ the LK model to characterize the chaotic dynamics. The parameters are the same as those set in Fig. \ref{fig3}, except $J=1.5J_{th}$, $\kappa =22$ $ns^{-1}$. As can be seen, the theoretical results agree well with the experimental data. The impact of the RF frequency is also investigated theoretically. Figure \ref{fig8}(b) shows the maxima and the minima of $C_{p}$ vary with the RF frequency in the range of 8 GHz. The minima of Cp remain almost unchanged within 8 GHz RF frequency and the maxima of $C_{p}$ show a raised area for intermediate RF frequency of 3-5 GHz. The raised area corresponds to the region near the relaxation oscillation frequency in Fig. \ref{fig2}(a) and is non-optimal for high entropy production.

\begin{figure}[htbp]
\centering
\includegraphics[width=0.85\linewidth]{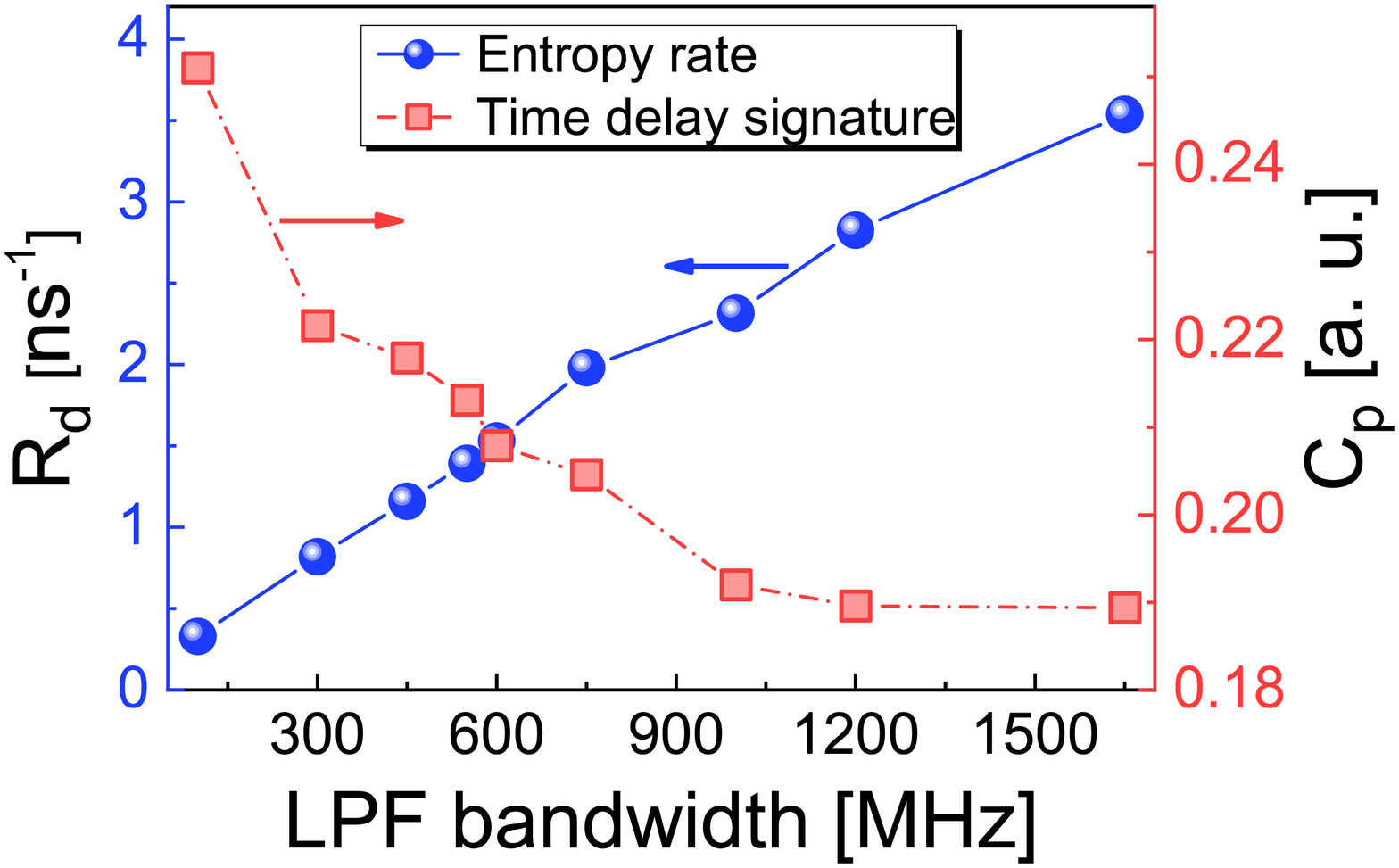}
\caption{Experimental results of the maximum $C_{p}$ in the TDS suppression and the corresponding $R$ of filtered chaotic laser as a function of LPF bandwidths.}
\label{fig9}
\end{figure}

The periodic crests and troughs of the $C_{p}$ in the TDS suppression are in accordance with $(2n+1)/(4\tau _{ext})$ and $n/(2\tau _{ext})$, respectively. The $R_{d}$ enhancement is consistent with the TDS suppression. It is well known that filtering chaotic signals does not influence the entropy rate in an ideal condition \cite{Badii88}, but in practice the entropy rate can be affected attributed to redundant or deterministic disturbances. Figure \ref{fig9} shows the measured maximum or crest values of $C_{p}$ in the TDS suppression and the corresponding $R_{d}$ of filtered chaotic laser under different LPF bandwidths. The original chaotic laser operates at the same parameters as in Fig. \ref{fig6}, except the RF frequency of 2.355 GHz.
As can be seen, the $R_{d}$ is enhanced as the LPF bandwidth increases, which coincides with the TDS suppression of the maximum $C_{p}$. The result exhibits that the LPF bandwidth can be optimized to redistribute the time series and reduce extra disturbances. The trough values of $C_{p}$ are about 0.015 that remain almost unchanged under different LPF bandwidths, and the TDS suppression ratio of 95\% is achieved compared to the original chaos. The power spectrum of the filtered chaos becomes flat and has a relatively broad effective bandwidth via the frequency-band extractor.

In conclusion, the entropy rate using the growth rate of average PE value in memory time is adopted to experimentally quantify and characterize entropy production of laser chaos and the physical randomness limit of shot noise. The PE difference of filtered laser chaos with varying embedding delay time is revealed experimentally and theoretically and high resolution maps of the PE difference for the laser chaos over the range of the injection-feedback parameter space is obtained. The PE growth of extracted shot noise and laser chaos with embedding dimension is faster than linear. In noise-dominated regime, the entropy rate can unambiguously discriminate between laser chaos and shot noise, compared to the indistinguishable PE difference between them. The inverse relationship between the entropy rate and TDS is observed under different injection currents and feedback strengths. By optimizing the RF frequency and LPF bandwidth of frequency-band extractor, 95\% TDS suppression ratio with the minimum $C_{p}=0.015$ is achieved compared to the original chaos, and the experimental results show good agreement with the theory. Thus, the technique offers a high-speed assessment of entropy rate of physical process and can be applied to regular, chaotic, and noisy signal evaluation, leading to potential applications in random number generation and secure communication.

\noindent\textbf{Acknowledgments}
This work was supported by the National Natural Science Foundation of China (NSFC) (61875147, 61671316, 61705160, 61775158, 61961136002), the Shanxi Scholarship Council of China (SXSCC) (2017-040), the Natural Science Foundation of Shanxi Province (201701D221116, 201801D221182), the Research Foundation-Flanders (FWO) (G0E7719N, G0G0319N), the Scientific and Technological Innovation Programs of Higher Education Institutions in Shanxi (STIP) (201802053), and the Program of State Key Laboratory of Quantum Optics and Quantum Optics Devices (KF201905).

\end{document}